%% file: graphene.tex
\documentclass[aps,prl,twocolumn,groupedaddress,floatfix,nofootinbib]{revtex4}

\usepackage{graphicx}
\usepackage{listings}
\usepackage{url}
\usepackage{multirow}
\begin{document}
\lstset{breaklines=true}

%\preprint{}

\title{Graphene Sails with Phased Array Optical Drive - Towards More Practical Interstellar Probes}


\author{Louis K. Scheffer}
%\email[]{SchefferL@janelia.hhmi.org}
%\thanks{}
\affiliation{Howard Hughes Medical Institute}


\date{\today}

\begin{abstract}
A spacecraft pushed by radiation has the major advantage that the power source is not included in the accelerated mass,
making it the preferred technique for reaching relativistic speeds.
There are two main technical challenges.  
First, to get significant acceleration, 
the sail must be both extremely light weight and capable of operating
at high intensities of the incident beam and the resulting high temperatures.
Second, the transmitter must emit high power beams through huge apertures, many kilometers in diameter, in order to focus
radiation on the sail across the long distances needed to achieve high final speeds.
Existing proposals for the sail use carbon or aluminum films.  Aluminum in particular is limited by a low
melting point, and both have low mechanical strength requiring either a distributed payload or complex rigging.
Instead, we propose here a graphene sail, which offers high absorption per unit weight, 
high temperature operation, and the mechanical strength to support simple rigging to a lumped mass payload.  
For the transmitter, existing proposals use a compact high power source, and focus the energy with a
large (hundreds to thousands of km) space-based lens.  For optical drive proposals in particular, existing proposals require launch
from the outer solar system, have severe pointing restrictions, and require difficult maneuvering of the beam source.
Here instead we propose
an active Fresnel lens operating at optical wavelengths, allowing smaller apertures of less mass, easier pointing with fewer restrictions,
and probe launch from the inner solar system.
The technologies for both the sail and the transmitter are already under development for other
reasons.
Worked examples, physically smaller and less massive than those suggested so far, 
range from a 1kg payload launched to 10\% of the speed of light by a transmitter only 25 times the mass of ISS, 
to a larger system that can launch a 1000 kg payload to 50\% of the speed of light.  
\end{abstract}

%\keywords{}

\maketitle

\section{Introduction}
A spacecraft pushed by radiation has the major advantage that the power source is not included in the accelerated mass.
This makes it one of the very few techniques that might achieve the velocities needed for interstellar probes with human-scale
travel times.
However, the technical problems associated with both the sail and the source are daunting.
The sail must be extremely light weight, yet either absorb or reflect photons well.  Since high intensities are
needed to achieve significant acceleration, it must operate at high temperatures.  Finally it should be mechanically
strong to simplify connecting it to the payload.

The transmitter has a very different, but similarly difficult, set of constraints.  
To focus radiation out to the required distance, it needs to be physically huge while generating coherent radiation
across the aperture.  It should be lightweight since it must operate in space and hence the components must be
launched and maneuvered into position.  It should allow a convenient place for launching probes, and allow
transmitting in different directions, to launch probes to nearby stars and later send the energy needed for returning
the data.
\section{Previous work}
A laser-driven aluminum sail was proposed by Forward in 1984\cite{forward1984roundtrip}.
The power source was a solar-powered laser, focussed by a 1000 km diameter Fresnel lens located at 15 AU from the sun, and
constructed of concentric rings of thin plastic film.
The system performance was limited by the melting point of the aluminum sail.  
The type of laser source was to be decided later as technogy advanced, and the details of aiming the laser
beam and keeping it focussed on the probe were not addressed.

A lightweight, microwave-driven probe was introduced as Starwisp\cite{forward1985starwisp}, with an aluminum mesh driven
by microwave radiation.  
The microwaves were to be generated by a solar power satellite, and used only incidentally for accelerating spacecraft.
The wavelength was not optimal for this application and required a truly enormous lens, much larger than the Earth.
Furthermore, the payload was tiny, only 4 grams, and had to be distributed across the sail in even smaller units, 
a technology not yet mastered and difficult even at a conceptual level.
Finally, later analysis, first informal \cite{scheffer1992melt} and then formal\cite{landis1999advanced}, 
determined the aluminum mesh would melt under the proposed illumination.  
Scaling the power density down to the point the mesh was not melted reduced
the available acceleration dramatically.  

All of these issues were addressed by Landis\cite{landis1999advanced}.
After recognizing that practical materials will operate in absorption and not reflection, 
he advocated a new figure of merit for potential sail materials.
For a sail dominated by absorption, the thrust is $P/c$, where $P$ is the power absorbed.  
This energy must be lost by radiation, which grows as $T^4$. 
Therefore the highest achievable acceleration scales as $T^4/m$, where $T$ is the operating temperature and $m$ the mass
per unit area, and the best performance is
obtained with lightweight materials with high melting points.  
Carbon films appeared the best choice in these studies.

Next, Landis proposed the use of shorter wavelength microwave illumination (3mm instead of 3cm), 
which reduced the lens size to 125 km.  
Finally, he used a larger, more practical, lumped payload of 80 grams.  
This required a complex fractal design for the rigging to connect it to the sail without overstressing any portion of the thin membrane.

Landis also briefly mentioned an optical sail made of sapphire.  
The sail is 57 nm thick to make it $\lambda/4$ at 400 nm incident light, to optimize reflectance.
It reflects 26\% of the light that falls on it while absorbing very little -
the absorption was too low to for him to measure experimentally, and he assumed 0.5\%.  
The system was only partially fleshed out, but has several numerical inconsistencies.
From the stated parameters (laser = 448 MW, reflection = 26\%, acceleration = 43.4 g), the
total mass of the sail + payload is 1.8 grams, which seems very small, particularly since there is a discussion
earlier in the paper of whether a payload can be reduced to 5 grams.  
The payload mass is not explicitly stated, but the text states the total mass is 2.3 times the sail mass, citing Forward's 1984 paper.
However, that paper used a ratio of 3 (``roughly one-third each of sail, structure, and payload''). 
Finally, the time of acceleration was 8.5 days, and the final velocity was 10\% c, but that combination gives 4.17 g, not 43.
Assuming that at least a Starwisp size payload is needed, 
we will assume what was meant was 4.17 g, 8.5 days of acceleration, 18 grams total mass, and 6 grams payload to 0.1c.
At the stated power density of 34 MW/m$^2$ this gives a sail 4.08 m in diameter.
Although the required power is indeed low, it requires a huge lens (2600 km in diameter) to focus on such a small sail out to the
required 75 AU.

An overview of these proposals is shown in Table \ref{table:summary}.

\begin{table}
\centering
\begin{tabular}{r | c | c | c | c}
Work & Starwisp & Landis 99 & Forward 84 & Landis 99\\
     & microwave & microwave & optical & optical\\ \hline
Power & 10 GW & 56 GW & 65 GW & 448 MW\\
Lens size & 50,000 km & 125 km & 1000 km & 2600 km\\
Wavelength       & 3 cm      & 3 mm   & 1 um & 400 nm\\
Sail size        & 1 km      & 100 m square & 3.6 km  & 4.08 m \\
Payload          &  4 gram   & 80 gram & 333 kg  & 6 gram\\
Acceleration     & 115 g     & 2.5 g  & 0.036 g & 4.17 g\\
Sail temp        & melts     & 2333$^\circ$ K & 600$^\circ$ K & 1583$^\circ$ K\\
Acc. Time        & 10 hours  & 18.5 hours     & 3 years & 8.5 days       \\
V at cutoff      & 0.1 c     & 0.005 c       &	0.11 c  & 0.1 c      \\
\end{tabular}
\caption{Summary of previous work}
\label{table:summary}
\end{table}
Graphene sails were mentioned briefly by Benford in \cite{benford2011starship}, where their high temperature capabilities were noted.  
This work also tried to minimize the cost of the transmitter, with different techniques than pursued here.
Matloff analyzed the use of graphene in solar sails\cite{matloff2012graphene}\cite{matloff2013speed}, noting the same overall advantages as mentioned here, 
though different criteria are needed for this application.  Bible et al.\cite{bible2013relativistic} also analyze a large scale phased array laser transmitter,
with many fewer but much larger elements than used here.  This work assumes a reflective, not absorbing, sail.
\section{Overview}

\begin{figure}
\begin{center}
\noindent
\includegraphics[height=6cm]{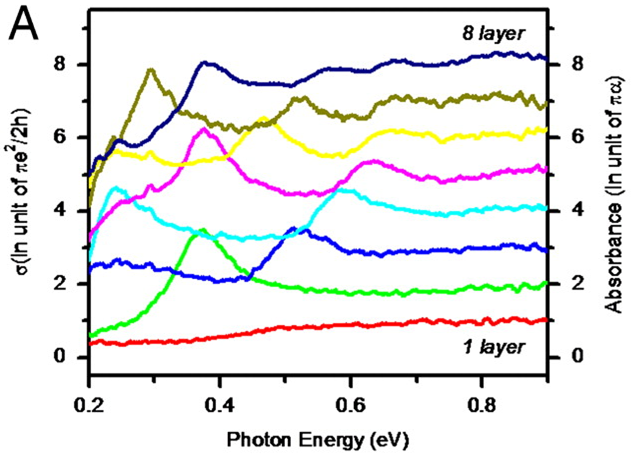}
\caption{Absorption as function of number of layers, from \protect\cite{mak2010evolution}.  Absorption in the visible is on the right edge (and futher right) of this figure.  Absorption is proportional to the number of layers, at least up to 8, and shows no sign of deviating from proportionality.}
\label{fig:absorb}
\end{center}
\end{figure}
This paper continues these trends to create a still more ``practical'' solution.
The first innovation is the use of graphene for the sail.  This provides three advantages over a carbon film.
The absorption is very high per unit thickness (in the optical, at least), it's mechanically very strong, 
which simplifies the rigging, and it's capable of high temperature operation.  
The second innovation is the use of a phased array, rather than a lens, for the optical drive.
This offers flexibility in aiming, probe launch from near-Earth, and does not require moving the source during operation.

Graphene consists of one or more layers, each a two-dimensional hexagonal array of carbon atoms. 
Since the studies of Landis, it has achieve considerable prominence.
Its invention garnered the Nobel prize in physics in 2010\cite{nobel2010}, and the award included a good summary of its properties.
Single layer graphene has a density of 0.77mg/m$^2$, and an optical absorbance of 2.3\%, 
independent of wavelength (in the visible spectrum).
Graphene is 0.345 nm thick, and has physical strength of 42 N/m, well above that of other materials of similar thickness.
The exact melting point is hard to determine experimentally\cite{steele2013non}, but appears to be about
4900$^\circ$ K\cite{zakharchenko2011melting}.

Consider a sail made of single layer graphene with a working temperature of 4000$^\circ$ K.  
It will radiate mostly in the optical, so its emissivity will be the same as the absorbance, or 2.3\%.  
Thus it radiates (2 sided) about $2\sigma\epsilon T^4$ or about 668 Kw per square meter.  
If it absorbs this all from one side, it applies a
force of 668 kw/c/m$^2$, or $2.2\cdot 10^{-3}$ newtons/m$^2$.  
Acting on the mass of $7.7\cdot 10^{-7}$ kg/m$^2$, 
this gives an acceleration of 2890 m/sec$^2$, or about 300 g.  
Naturally the addition of a payload, and the limits of available
power, will reduce this figure considerably, but it shows the potential.

Furthermore, a graphene sail is strong enough to use simple rigging to propel a large, dense object, which 
is not possible with Starwisp.  
Starwisp was constructed of 0.1 micron wires on .3 mm spacing.  Assuming
high strength aluminum of 440 MPascal strength, this gives a strength of 0.0012 n/m.  
Under the proposed illumination of 8.6 kw/m$^2$, perimeter
rigging (like a parachute) would exceed the tensile strength of the sail at a radius of 100 m, with a total force of 0.7 newtons.
Graphene, at over 10,000 times the tensile strength of the aluminum grid, allows sails of almost any size to be rigged only at the rim.

Graphene comes in both single and multi-layer forms, and the use of multi-layer graphene helps considerably.  
Each additional layer absorbs 2.3\% of the remaining light, so for small number of layers the absorption is nearly
proportional to the number of layers\cite{nair2008fine}.  
This has been measured directly for up to 8 layer graphene (see Fig \ref{fig:absorb}, from \cite{mak2010evolution}), where the property holds without
any sign of converting to bulk graphite, in the near-infrared (1 ev) and shorter visible wavelengths

As additional layers of graphene are added, the mass goes up, but so does the absorption (and hence thrust from constant flux).  
The temperature is unchanged since emissivity rises at the same rate as absorption.  
For a graphene sail alone, a single layer is optimal, since adding another layer doubles the mass, but does not quite double the absorption.
But if the payload is a significant fraction of the total, then
the overall performance goes up as the number of layers increase.  

The limit to this process is when each additional
layer adds less absorption than it does mass, 
or when the properties of multi-layer graphene change to those of bulk graphite\cite{mak2010evolution}.
Assuming the trend in Fig \ref{fig:absorb} continues to many layers, and the sail mass is equal to the payload (a reasonable approximation, as shown later), then the optimum number of layers is 54, for a thickness of 18.6 nm and absorption of 72\%.  
At this thickness an additional layer contributes 0.92\% to both absorption and mass.
\subsection{Comparison of sail materials}
\begin{table}
\centering
\begin{tabular}{c| r | c | c | c }
(a)&Material & Aluminum & Sapphire & Graphene \\ \hline
(b)&Thickness & 16 nm & 57 nm & 17 nm \\
(c)&Mass/m$^2$ & 43 mg  & 226 mg  & 42 mg \\
(d)&Reflectance& 82\% & 26\% & 0\% \\
(e)&Absorbance & 13.5\% & $1.7\cdot 10^{-7}$ & 72\% \\
(e)&Thrust/W,  & 1.77/c & 0.52/c & 0.72/c \\
(f)&Emissivity & 0.06 & 0.10 & 0.72 \\
(g)&Melting pt & 933$^\circ$ K     & 2303$^\circ$ K     & 4900$^\circ$ K \\ \hline
(h)&Area for 333 kg      & 7.7 km$^2$& 1.4 km$^2$& 7.9 km$^2$\\
(i)&Radius for 333 kg    & 1570 m & 685 m & 1589 m\\
(j)&Can radiate/m$^2$    & 1.02 KW & 63 KW & 9.3 MW\\
(k)&Incident power/m$^2$ & 7.54 KW& 136 MW& 12.9 MW\\
(l)&Power in beam        & 58.4 GW & 201 TW & 102 TW\\
(m)& Thrust              & 346 N  & 348 KN & 246 KN \\
(n)&Total mass           & 1000 kg  & 1000 kg  & 700 kg\\
(o)&Acceleration         & 0.35 m/s$^2$ & 348 m/s$^2$ & 351 m/s$^2$ \\
(p)&Time to 0.1c         & 2.73 yrs& 1.00 day& 0.99 day\\
(q)&Distance to 0.1c     & 8690 AU & 8.65 AU& 8.56 AU\\
(r)&Total energy (J)& $5.07\cdot 10^{18}$ & $1.73\cdot 10^{19}$ & $8.74\cdot 10^{18}$\\
(s)&Transmitter radius   & 253 km & 576 m & 246 m \\
\end{tabular}
\caption{Comparison of sail material options.  Data for the aluminum sail from \protect\cite{forward1984roundtrip}.  Data on the sapphire sail from \protect\cite{landis1999advanced}, except for transparency which is estimated from data in \protect\cite{dobrovinskaya2009properties}.  Below the line are computed properties for the case of a 333 kg payload to 0.1c .}
\label{table:compare}
\end{table}
The three materials proposed for optical sails are aluminum, sapphire, and graphene.
Their properties are compared in Table \ref{table:compare}.
One way to compare them is to examine the results of using each material to accelerate a canonical 333 kg payload to 0.1c,
an example already used by Forward.  We will start by assuming an optimal transmitter for each sail.

The calculation proceeds as follows.  
From both Forward and Landis, and later in this work, the sail mass should roughly equal the payload mass.  
So the first step is to find the size of a 333 kg sail of each material, entries (h) and (i) in the table.  
Next, using the emissivity and assuming 2-sided radiation, figure out how much power the
sail can radiate, per m$^2$, without exceeding 2/3 of melting temperature.
This is shown in line (j).
The difference is dramatic, driven by the 1770-fold more power that can be radiated
at the higher temperature with the higher emissivity.  
This difference in absorbed power dominates all the aluminum-graphene differences that follow.

The next step is to use the absorption to figure out the incident power the sail can support.
For sapphire, this gives an enormous value due to its very low absorption.
However using the full value tolerable by the sail would doubtless overheat the payload. 
Even for the graphene sail, the illumination is already 10000 suns, requiring a roughly 99.99\% reflective payload to avoid overheating. 
We will give the sapphire sail the benefit of the doubt, and assume that a 99.999\% reflective mirror over the required 10\%
bandwidth is possible.\footnote{The 10\% bandwidth requirement is caused by redshift as the sail accelerates.  In theory, this could also be
dealt with by changing the laser wavelength as the probe accelerates, to keep the wavelength constant at the probe.  Then the mirror needs only be 99.999\% reflective at one wavelength, which has been done.
But a super high power, efficient, and tunable laser seems at least as much of a stretch as a high bandwidth mirror.}  This limits the beam intensity to about 100,000 suns, or $1.36\cdot 10^8$ W/m$^2$.
Results are shown in line (k).

Given the incident intensity, the area, and the efficiency, the thrust can be computed, line (m). 
This acts on the mass, and here the lesser strength of aluminum and sapphire come into account.
We assume the mass of the supporting structure is equal to the payload mass for aluminum (as assumed by Forward and Landis). 
We also assume this for sapphire, though this again might give it the benefit of the doubt since
the huge intensity in that case would likely require sapphire rigging.
And for graphene, its strength allows simpler rigging (we will show this later), so we assign it 10\% of the mass of the sail.
The total mass is shown in line (n).

At this point we have the acceleration (o), from which we compute the time (p) and distance (q) needed to reach 0.1c.
From the distance, and the size of the sail, we derive the radius of the transmitter needed (s)
(assuming $\lambda = 500$ nm in all cases, though this too is an approximation).  
From the time of acceleration, and the beam power, the total energy expended (r) can be computed.

As mentioned earlier, the dominant factor in these calculations in the melting point, since it is raised to the 4th power.
The aluminum sail has the best efficiency - it generates more thrust per
photon and has comparable mass to graphene (and much less than sapphire).  However, it has a low maximum temperature, limiting
the peak acceleration, hence requiring a long distance, a long time (almost three years),  
and a proportionally large lens - 250 km in radius in this case.  (This is half the radius found in Forward since we
are using half the wavelength, but otherwise the calculations largely agree.)
In return for the slow acceleration, the source power is much smaller, ``only'' 58 GW.

The sapphire sail is an intermediate case.  
It has the highest total energy needed, and the highest power source, due to its low efficiency.
But the acceleration is high due to the enormous intensity it can support, so the acceleration distance is short, comparable
to graphene.  
But the lens is bigger, twice the size of graphene, since the higher mass per unit area means a smaller sail.  
Finally, two other factors would need to be taken into account for a sapphire sail.
The reflectivity is tuned to light of a specific wavelength arriving perpendicular to the surface, and will work less well as redshift increases
or if the sail is not at right angles to the incoming radiation.

The graphene sail's combination of strong absorption, high temperature operation, and light weight yields the smallest lens, 
as small as 246 m in radius.  (In practice, other considerations such as energy density may impose larger minimums)
The source power is high, 102 TW, but allows a probe launch once per day, rather than once every three years.

The possibility of a smaller lens allows the consideration of more complex lens architectures.
Since the lens used by Forward was large, it was essential to keep the mass per unit area as low as possible, 
and hence it was a passive structure only a half wavelength thick on the average.  
But if the lens size can be reduced, perhaps by an order of magnitude, then a much thicker and more complex
lens can be considered, and still have less mass than the large thin lens.

As a side note, if you want only to accelerate the sail, and are not concerned about the payload, 
then sapphire offers some incredible performance.
The absorption of visible light by sapphire is extremely low.
Even at high temperatures (1500$^\circ$ K) the absorption is only about 0.03 per cm.\cite{dobrovinskaya2009properties}
So a 57 nm film should only absorb about $1.7\cdot 10^{-7}$ of the incident light.  
This would allow an enormous incident intensity of 411 GW/m$^2$.
Reflected with 26\% efficiency, this generates a force of about 712 N/m$^2$, which acting on a mass of 226 mg/m$^2$, creates an acceleration of
$2.68\cdot 10^6$ m/s$^2$, or about 273,000 g.  If this acceleration could be sustained, it would reach the speed of light in less than
two minutes.  
Unfortunately, using this theoretical performance for anything more than accelerating the sail alone seems impractical.  
Any payload attached to the sail
will be instantly vaporized unless it is either as transparent as the sapphire or far more reflective than any known material.
\begin{figure}
\begin{center}
\noindent
\includegraphics[height=5cm]{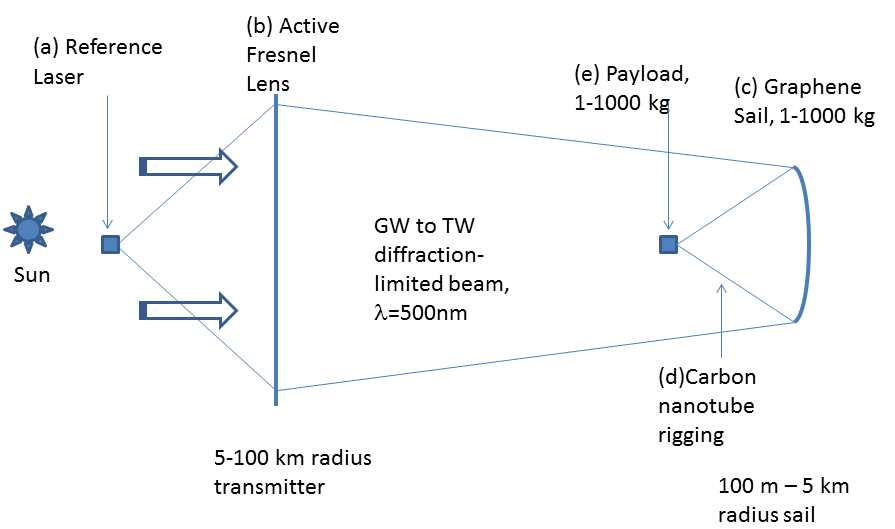}
\caption{Schematic diagram of system.}
\label{fig:Year}
\end{center}
\end{figure}
\section{Overall architecture}
Existing proposals require two large space-based structures - beam generation and the lens.  For example, the proposal of Forward
starts with a 65 GW solar powered laser.  At 25\% efficiency and in Earth's orbit, this involves 191 km$^2$ of collecting area.
This generates the beam which is then focussed by the 1000 km diameter lens in the outer solar system.

Given the smaller lenses that are possible with graphene based sails, it is perhaps possible to combine these two
functions into one structure, one that receives incoherent raw sunlight on one side and emits a coherent optical
beam on the other, based on an optical reference signal.  
This is in essence a Fresnel lens, using a thin flat plate with local modification of phase to achieve lens operation.  
It is active since it adds power to the reference signal, and since the local phase changes are under computer control.
This allows relaxing the physical constraints on the lens, varying the focal length during operation, and control of
the beam profile.

Such a system might be built as shown in Fig \ref{fig:Year}.  
A phase reference laser at (a) illuminates the back of the a multi-km radius transmitter, (b).  
This transmitter uses sunlight (thick arrows), solar cells, 
and phase controllable semi-conductor lasers to create a multi-GW coherent optical beam at optical wavelengths.
This is partially absorbed be the graphene sail at (c), which pulls the lumped payload (e) through the rigging (d).
Since the rigging is at the edge, but the thrust is distributed, the sail becomes convex which is needed for
passive centering of the sail on the transmitted beam\cite{singh2000characterization}.
The acceleration of the sail and payload continues until the transmitter can no longer focus a significant fraction of
its power on the sail.  

This approach solves several problems inherent to the configuration with a separate power source and passive lens.
First, to keep the mechanical tolerances of the lens reasonable, it
was built with an extremely long focal length - many AU for a 1000 km diameter lens.  
In the paper, the lens was proposed to be located 15 AU from the sun.  
At this distance, the lens-transmitter combination can only cover a very small part of the sky, within a few degrees of the sun-lens line.
For the specified 15 AU distance, this line will orbit the sun each 58 years.
This could require additional lenses should it be necessary to send a similar beam to the same target star 
at a particular time many years later, perhaps to power data return.

The next problem is that the probe(s) must be launched from the outer solar system, 
presumably delivered there by some more conventional means.  
This is closely related to the third problem, that the laser source needs to be located at one specific location
in space.
As with any conventional lens, as the probe moves away, the needed location for the source moves closer to the lens.
This results in a tradeoff, where a larger focal length lens gives less required movement of the source, but the closest
attainable focus is further away.

For example, the design might want the laser to have to move at most 1 AU as the probe accelerates, as
might be needed if the laser is solar powered and must remain close to the sun.
Assuming a 15 AU lens location, this
requires a 14 AU lens focal length which can only focus as close as 210 AU from the lens, or 225 AU from the sun.
This is much further out than any man-made object to date (Voyager is at 131 AU as of 2015, after almost 40 years of flight.)
Even with this large starting distance,  the light source must accelerate up to 7 km/sec, then decelerate, during the launch of the probe.  
A shorter focal length lens can offer the probe a starting spot closer to the sun, but requires more movement and higher speeds from the laser.
If you want to start at only 30 AU from the sun, for example, the focal length needed is 7.5 AU, and the focal point
moves more than 7 AU (making solar power difficult) at a peak speed of 270 km/sec (well above the Sun's escape velocity).

An active lens solves all these problems.  
Since the active lens focal length is under computer control and can be changed at will, the lens could be built at a Venus orbit (for example) and 
the probes constructed and launched near Earth.  The lens can cover a wide portion of the sky - up to a 45 degree offset from the
sun-lens line can be accomodated with a $\sqrt{2}$ degradation in power.  
In the worst case, 8 such lenses, four in the orbit of the ecliptic and four in sun polar orbit, could cover the whole sky at any time.

The following sections work through some of the needed calculations, then size and performance estimates are
given for a variety of possible interstellar missions.

\section{Focussing distance}
How far away can a transmitter of radius $R$ focus a beam onto a sail of radius $r$, using radiation of wavelength $\lambda$?  
As in \cite{forward1984roundtrip}, we define this distance as that where the sail just fills the main lobe of the 
diffraction limited circular beam.  At this point 84\% of the total beam energy falls upon the sail.
If the transmitter is radius $R$, the sail radius $r$, then this happens at distance
$$d = \frac{4 r R }{2.44 \lambda}$$
This distance is known as ``cutoff'' in the literature.

After cutoff, the final speed can still be increased, but
the acceleration now falls as $1/s^2$, where $s$ is the distance.  
In the classical case, if continued forever, the energy will eventually double and the 
speed increased by a further $\sqrt{2}$.  This result is stated, but not derived, in \cite{forward1985starwisp}.
Perhaps the simplest derivation is via calculating the work done on the craft by the photons.
Assume we have force $F_0$ applied over a distance $S_0$, with force $F_0 (S_0/s)^2$ thereafter.
Then the energy imparted up to cutoff is simply $F_0 S_0$, and the work up distance $s$, greater than cutoff, is
$$E(s) = F_0 S_0 + \int_{S_0}^s F_0 (S_0/s)^2 ds$$
Integrate to get
$$E(s) = F_0 S_0 (2 - S_0/s)$$
and the energy is at most doubled by acceleration after cutoff.
However, acceleration after cutoff is an inefficient use of a very expensive beam generator, 
and hence for the purpose of this paper we assume drive stops when the beam can no longer focus down to the size of the sail.

It should be noted that there is confusion on this point in the literature.  
Although Forward correctly states (see \cite{forward1985starwisp}, equation (12)) 
that it is {\it energy} that can be doubled by post-cutoff acceleration, 
many of the examples in \cite{forward1985starwisp}, \cite{landis1999advanced}, and \cite{forward1984roundtrip} 
assume instead it is the {\it velocity} that is doubled.  
This results in some unjustifiably optimistic final velocities.
\section{Relativistic calculations}
Since the whole point of this operation is to attain relativistic speeds, relativistic calculations are necessary.
Consider a mass $M_0$ propelled by radiant power $P_0$, which is entirely absorbed.  This induces a force $F_0 = P_0/c$ and
and initial acceleration $A_0 = \frac{P_0}{M_0 c}$.  
Now as the mass picks up speed, the acceleration goes down in two ways.
First the incoming photons are red-shifted, so the force goes down
$$ F(v) = \frac{P_0}{c} \sqrt{\frac{1-v/c}{1+v/c}}$$
In addition, the mass goes up
$$ M(v) = \frac{M_0}{\sqrt{1 - v^2/c^2}} $$
and hence the acceleration (in the rest frame) is
$$ A(v) = \frac{P_0}{M_0 c} \sqrt{\frac{1-v/c}{1+v/c}} \sqrt{1 - v^2/c^2}  = \frac{P_0}{M_0 c} (1-v/c)$$
This sets up a differential equation, and assuming $a(0) = v(0) = s(0) = 0$, the solutions are:
\begin{equation} a(t) = A_0 e^{\frac{-A_0}{c} t} \end{equation}
\begin{equation} v(t) = c (1 - e^{\frac{-A_0}{c} t})  \end{equation}
\begin{equation} s(t) = c t - \frac{c^2}{A_0}(1-e^{\frac{-A_0}{c} t})  \end{equation}
Since the distance D to which the beam can focus is unaffected by relativistic corrections, to find the final relativistic speed we invert (3), setting $s(t) = D$, to find the time $t$ of last drive.  We plug this time $t$ into equation (2) to get the final speed.

\section{Disk stiffness}
To keep the sail in the beam without active control, it needs to be convex as seen by the beam\cite{singh2000characterization}.  
This could be achieved by attaching the rigging to the rim of the disc, then rotating the disk to keep it from collapsing.

A graphene disk can rotate at high speeds.  A uniform disk of density $\rho$, radius $R$, and Poisson's ratio $\nu$,
rotating with angular speed $\omega$ has a maximum stress $\sigma$ \cite{hearn1997mechanics}
$$ \sigma_M = (3+\nu)\frac{\rho\omega^2 R^2}{8}$$
Working backwards from the properties of graphene ($\sigma = 10^{11}$ Pascal, density = 1300 kg/m$^3$, Poisson's ratio of 0.25), 
we find (for example) that a disk of 1km radius can rotate with a speed of 16 radians/sec.  
This gives a speed at the edge of 16,000 m/s, and an acceleration of $256,000$ m/s$^2$, or about 25,000 gs.
This is much larger than the linear acceleration and keeps the disk from collapsing.

The hoop stress at the outer edge is given by:
$$ \sigma_H = (1 - \nu) \frac{\rho\omega^2 R^2}{4}$$
or, comparing to the maximum stress above
$$ \sigma_H = \sigma_M \frac{(1-\nu)}{(3+\nu)}\frac{8}{4}$$
Using the above values, the maximum hoop stress is therefore about $4.6\cdot 10^{11}$ Pascal.

Assuming the rigging is at a 45$^\circ$ angle, the sail will experience a force trying to collapse the sail that is equal to the thrust. 
The outward hoop stress from the spinning sail needs to exceed this force to prevent the sail from collapsing.  
The calculations below find the rim size, the distance from the edge where the integrated outward stress exceeds the inward stress from the rigging, assuming the disc is rotating as fast as possible.
In all cases this is a very small fraction, much less than 1\%, of the sail radius.
In practice, the sail would not rotate at its maximum possible rate, but 
even so the rigging only needs to attach to a small area near the rim, and the rest of the sail will be in pure tension.
\section{Optimum sail size}
This section demonstrates, that at least in the classical case, the optimum final velocity (for constant thickness) is achieved when the sail has the same mass as the payload.
Changing the sail size has two effects - it changes the total accelerated mass, and it changes the distance to which the transmitter can focus.  
It does not change the thrust since the entire beam is incident upon the sail independent of the sail size.  So we have
$$v(r) = \sqrt{2 a(r) s(r)}$$
If we set $P$ to be the payload mass, $E$ the absorbed power, and $M_0$ the mass per unit area times $\pi$, then we get
$$ v(r) = \sqrt{2\cdot  \frac{E}{c (P + M_0 r^2)} \cdot \frac{4 r R}{2.44 \lambda}} $$
Now we need to find the optimum $r$.  First, since the square root of a positive number is a monotonic function, we can just find the optimum of 
$$ 2\cdot  \frac{E}{c (P + M_0 r^2)} \cdot \frac{4 r R}{2.44 \lambda} $$
next, $E$, $R$, $c$, and $\lambda$ are just multiplicitive constants, and do not change the optimum $r$.  So we just need to find the $r$ that maximizes
$$ \frac{r}{P+M_0 r^2}$$
The derivative with respect to $r$ is
$$ \frac{(P+M_0 r^2) \cdot 1 - r \cdot 2 M_0 r}{(P+M_0 r^2)^2}$$
This can only be 0 if the numerator is 0, so we set
$$P + M_0 r^2 - 2 M_0 r^2 = 0$$
This is 0 when $M_0 r^2 = P$, or in other words when the mass of the sail equals the mass of the payload.  
From the numerical evidence, the same appears true in the relativistic case, where $v = \sqrt{2 a s}$ is replaced by inverting equation 3 and then inserting the time into equation 2.  The analytic proof of whether this is precisely true is left to the interested reader.

\section{The active lens}
How can an active lens such as proposed here be constructed and aimed?  
Unlike the sail, the active lens had many possibilites for its construction.
The arguments here are therefore to show that it is plausible that it could be built, not a blueprint for construction.
\begin{figure}
\begin{center}
\noindent
\includegraphics[height=7cm]{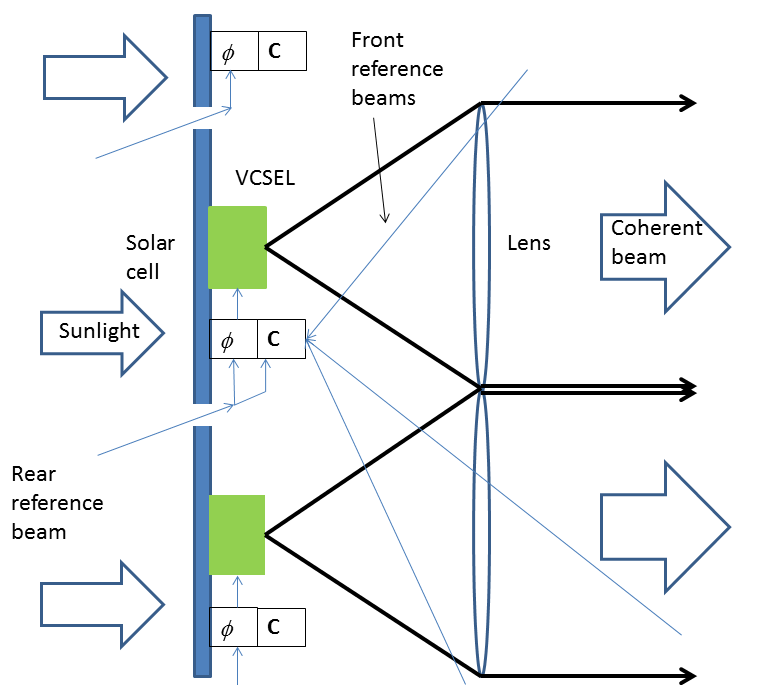}
\caption{Detail view of active lens. Module C computes the required phase shift based on the measured location of the pixel and the desired beam direction.  Module $\phi$ phase shifts the reference beam, which is then amplified by (or locked to) the VCSEL.  The VCSEL gets its power from the solar cell on the sunward side.  The distance to the lens is not to scale, as the exit divergence of VCSELs is 5-10$^\circ$. }
\label{fig:detail}
\end{center}
\end{figure}
The main idea is to combine a high efficiency solar cell\cite{fr2013world} with a high efficiency laser\cite{li2007near}.
To minimize the size, mass, and cost of the lens, the power output per unit area should be as high as practical.
This implies high illumination on the solar cells, either by locating the lens closer to the sun than the Earth, or perhaps using
large thin reflectors to concentrate additional light onto the array.
The limiting factor is when the temperature of the array becomes too high for reliable operation.
We pick a reference insolation of 4 kw/m$^2$.  
Since good solar cells convert $>$40\% of light into electricity, and good lasers have more than $>$60\%
efficiency, this gives an overall efficiency of perhaps 25\%, generating 1kw/m$^2$ of drive power.
If we assume that of the 4kw/m$^2$ incident, 1kw/m$^2$ is reflected and 1kw/m$^2$ is the output beam, then this leaves 2 kw/m$^2$ which
must be re-radiated as waste heat.  If the lens is also an efficient radiator in the IR,  this leads to a temperature of 91$^\circ$ C.  

The array consists of pixels spaced a few hundred wavelengths apart.  
Each pixel continually adjusts its phase to generate a coherent output beam, based upon its physical location and the desired output direction.
The reference beam arrives from the rear.  
It is phase shifted on a per-pixel basis, then amplified and emitted.  
Since the pixels are many wavelengths apart (perhaps 200 wavelengths, for a 100 micron spacing and 500 nm light), 
each pixel must emit only over a narrow angle (roughly 1 over the spacing in wavelengths) to avoid serious grating losses.
This means the disk must be physically aimed close to the required direction.

Multi-junction III-V compound cells are currently favored for their high efficiency, and small thickness compared to silicon cells\cite{geisz200840}.
Assuming that the lens is dominated by the mass of the 10 micron thick solar cells with the density of GaAs, the mass is about 52,000 kg/km$^2$.  
The International Space Station, for comparison, is about 420,000 kg, or equivalent to about 10 km$^2$ of transmitter.
It is possible that the mass of the lens might be reduced further, since thinner solar cells are in active development, though none yet have comparable efficiency.
Thin film cells made of CdTe can have active areas as thin as 1 micron\cite{shah1999photovoltaic},
and there is research into cells just a few atoms thick, using the same high optical
absorption exhibited by graphene\cite{bernardi2013extraordinary}.

\subsection{Laser}

VCSELs (Vertical Cavity Surface Emitting Lasers) are a likely candidate for the laser amplifier.  They can be built with the required efficiency, but require a lens since their native beam width is too high.

\subsection{Aiming the active lens}

\begin{figure}
\begin{center}
\noindent
\includegraphics[height=6cm]{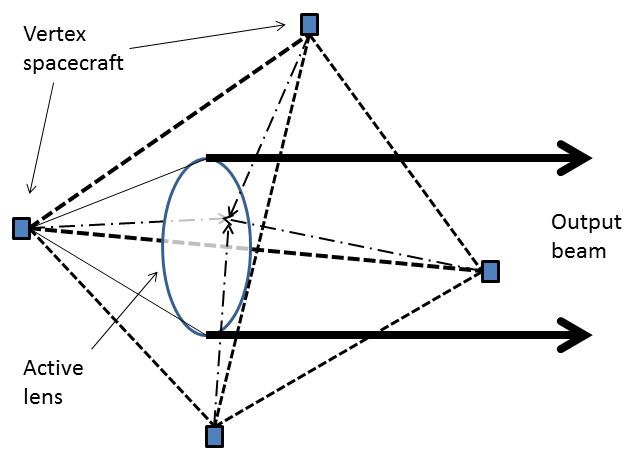}
\caption{Aiming the lens - the big picture.  The four filled squares fly in formation, defining a reference tetrahedron by keeping constant length on the six thick dashed lines. Each element of the lens computes its coordinates by comparing the modulation and phase of illumination from the 4 tetrahedron defining satellites, as shown by the thin dash-dot lines.}
\label{fig:tetra}
\end{center}
\end{figure}
Each pixel of the lens must know its location relative to the reference source and desired beam direction, in order to compute the correct phase.
The coordinate system is defined by four spacecraft flying in formation, arranged in a tetrahedron a few hundred km on a side.  
Within the tetrahedron, the transmitter disk rotates, keeping it roughly flat.  Each of the vertex spacecraft
illuminates the transmitter disk.  Each point on the disk compares the phase of the illumination, and its modulation,
to determine its position.   It then applies the appropriate phase correction to the reference signal, resulting
in a single coherent output beam.  
The phase correction applied will vary throughout the mission, both because the focal length is changing, 
and since the goal is not a single sharp focus, but a uniform intensity across the sail.

The vertex spacecraft will need station-keeping.  In particular, the vertex generating the reference beam needs to be almost
directly in line with with the desired output beam and the center of the rotating lens.
This is needed to keep the phase updates down to a reasonable rate.  
The other vertex spacecraft must at least measure, and potentially control, their location relative to the other vertices to within a fraction of a wavelength.

The steering of the vertex spacecraft can potentially be done with solar sailing forces, 
as discussed later for the disk, or with tiny thrusters as developed for the 
LISA (Laser Interferometer Space Antenna)\cite{danzmann2003lisa} mission.  
The measurement of the relative positions might use a combination of the millimeter wave distance
measurement developed for GRACE\cite{tapley2004gravity} and GRAIL\cite{enzer2010grail} and their follow-on missions\cite{sheard2012intersatellite}, 
and the extremely precise optical measurement technology developed for LISA.

\begin{figure}
\begin{center}
\noindent
\includegraphics[height=3cm]{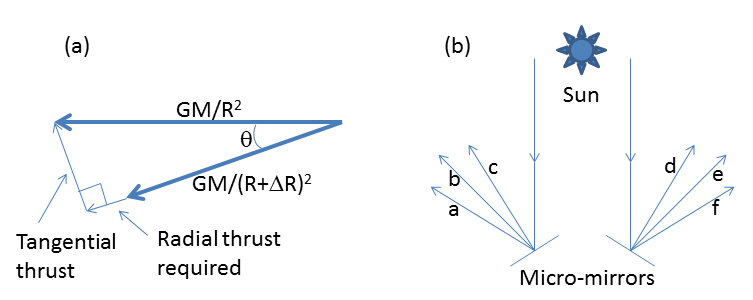}
\caption{Stationkeeping for formation flying.  (A) is the balance of forces needed to say in the same relative position when at an angle $\theta$ from the central element, and at distance $\Delta R$ radially.  (B) shows how reflective controls with micro-mirrors might work.}
\label{fig:aim}
\end{center}
\end{figure}
The lens is a thin membrane and has very little structural stiffness, though it will probably spin slowly to help keep it flat.
Therefore we model it as an independent group of membrane elements that must fly in formation.  
In addition, the beam source (behind the lens) and the metrology elements (beyond the lens) must also fly in formation with
the lens itself.
What are the forces needed to do this, and how can they be generated?

The main force that must be countered is the Sun's gravity.   
This vector always points towards the Sun, and formation flying requires adding additional forces to make the vector force on each element equal to the force on the central element of the array.

Consider an element that is at an angle $\theta$ to the central element, as seen from the star.  
Let the distance from the central element to the star be $R$, and the additional distance for the element under consideration by $\Delta R$.  Now from the balance of forces as shown in Fig \ref{fig:aim}, we need a sideways of $GM/R^2\cdot \sin(\theta)$, and an additional radial force of $GM(1/R^2\cdot \cos(\theta) - 1/(R + \Delta R)^2$.

\begin{figure}
\begin{center}
\noindent
\includegraphics[height=8cm]{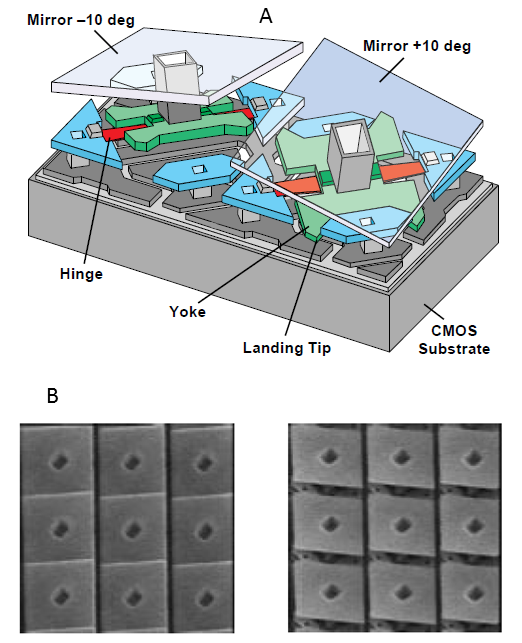}
\caption{Micro-mirror array from Texas Instruments, from \protect\cite{hornbeck1997digital}.  (A) shows the operation of two pixels.  (B) shows a portion of the roughly 1000 x 1000 array, with the pixels deflected one way in one image and the other way in the second image.  Modern devices feature a $\pm 12^\circ$ degree deflection.}
\label{fig:dmd}
\end{center}
\end{figure}
To get the required 4 kW on the array, it must be 0.58 AU from the sun.
If we wish to control arrays of up to 100 km in radius, then $\theta < 1.15\cdot 10^{-6}$ radians.
Assuming a maximum $45^\circ$ tilt from the normal to the sun, then $\Delta R < 70.7$ km.
This implies a radial acceleration needed of $2.82 \cdot 10^{-8}$ m/sec$^2$, and a perpendicular acceleration of $2 \cdot 10^{-8}$ m/sec$^2$.

\begin{table*}
\centering
\begin{tabular}{ r c c c c c}
\multirow{2}{*}{Example} & \multicolumn{1}{|c|}{1 kg}    & \multicolumn{1}{c|}{1000 kg} & 4 kg    &\multicolumn{1}{|c|}{1000 kg} & limited\\
                         & \multicolumn{1}{|c|}{to 0.1c} & \multicolumn{1}{c|}{to 0.15c}& to 0.5c &\multicolumn{1}{|c|}{to 0.5c} & layers \\ \hline
\input{tab}
\caption{Worked examples of different performance levels}
\label{table:1}
\end{table*}
Consider now a 1 meter square of material.  
At 10 microns in thickness, and a density of 5200 kg/m$^3$, there is about 52 g of material.  
The incident solar energy is about 4 kw, giving a force of 4000/c = $1.33 \cdot 10^{-5}$ Newtons.
If we can convert this solar radiation force into steering thrust, we have plenty of authority for station keeping.  
We can do this by covering a small portion of the sunward side of the lens with steerable reflector arrays, 
such as the TI micro-mirrors\cite{hornbeck1997digital} shown in figure \ref{fig:dmd}.   

One possible configuration could feature pairs of mirrors, as shown in Fig 3(b).   
The normal un-deflected positon would result in reflections \textbf{b} and \textbf{e}.  
Switching to \textbf{c} and \textbf{d} would increase the radial thrust, 
while switching to \textbf{a} and \textbf{f} would decrease it.
Similarly, switching to \textbf{a} and \textbf{d} would generate a rightward thrust, while leaving the radial thrust largely unchanged.
Reflections \textbf{c} and \textbf{f} would generate a leftward component.
Other configurations will generate a mixture of radial and sideways forces.
Since each chip consists of millions of mirrors, and each mirror can be switched with microsecond speed, fine control of
forces should be straightforward.

Assuming at $\pm 12^\circ$ deflection from a static angle of $45^\circ$, then covering only $10^{-3}$ of the sunward surface
with micromirror arrays will generate sufficient thrust for stationkeeping.
Since these actuators require no power to remain in position, and cover only a small fraction of the surface,  
their effect on the mass and power budget should be small.
\subsection{Worked Examples}
Performance of the system is very dependent on the size of the transmitter.
Increasing the radius $r$ of the transmitter helps in two ways.  
As long as the temperature limit is not reached, then acceleration scales as $r^2$, 
plus the distance to which the beam can focus scales as $r$.  
Since, classically, $v = \sqrt{2as}$, the final speed obtained scales as $r^{3/2}$.  
So a little less than 5-fold increase in radius can increase the final speed by a factor of 10.   
We will therefore consider a ``small'' transmitter with a radius of 7.65 km, a ``medium'' system with a radius of 32 km,
and a ``large'' system with a radius of 80 km.
Assuming the mass of the transmitter is dominated by the mass of existing GaAs solar cells, 
this results in 25, 400, and 2550 times the mass of the International Space Station (ISS).
For comparison, the passive 1000 km diameter sail of Forward was 560,000 tons, or 1333 times the mass of ISS.

The exact drive wavelength is not critical, though it should be far enough into the visible so that even when redshifted,
it still remains in the region of high absorption as shown in Fig \ref{fig:absorb}.  This requires the arriving photon to have
an energy of at least 0.7 ev, and if the maximum speed obtained is $0.7c$, then the source photon should have a wavelength of
745 nm or less.  Here we assume a driving wavelength of 500 nm.

First consider the small transmitter, pushing a 1 kg payload.  
This beam drives a 87.5m radius sail (also about 1 kg), composed of 54 layer graphene. 
The sail absorbs 72\% of the incident light, reaching a temperature of 2865$^\circ$ K and generating 438 N of thrust.  

This thrust is conveyed to the 1kg payload by carbon nanotube rigging (d).
The thrust acting on the payload, sail, and rigging gives an acceleration of 219 m/s$^2$, or 22 gs.  A 7.65 km aperture at
500 nm can focus down to a 87m radius out to a distance of 14.7 AU.  
This allows the probe to accelerate for 1.6 days, reaching a speed of 0.103c.
Relativistic corrections are small for such ``slow'' velocities, and including them gives a final speed of 0.100c.

A bigger transmitter allows larger payloads, higher speeds, or both.
Consider scaling the transmitter up to the medium size with a 32 km radius, still 3 times less massive than the
passive lens in \cite{forward1984roundtrip}.
The first thought might be to drive the same 1 kg payload and sail with this larger transmitter, but that melts the sail.
To keep the temperature down, we need at least a 4x bigger sail (and even that gives 4140$^\circ$ K).
This accelerates the 4 kg payload at a little less than 100 g.
A classical calculation gives a final speed of 0.625c and hence is clearly not appropriate.  
The relativistic calculation of section III shows a final speed of 0.5c.

Alternatively, also with the medium size transmitter, we can launch a larger payload more slowly.  
A 1000 kg payload, with a correspondingly larger sail, accelerates at only 0.4 g.
However, we can keep this up for 146 days, leading to a final velocity of 0.149c after relativistic correction.  The sail in this case is cool
at less than 1100$^\circ$ K.

The fourth example demonstrates that we can use an even larger transmitter to send big payloads fast.  
An 80 km radius transmitter can send a 1000 kg payload to
50\% of the speed of light in 100 days at 2.5 g.

The final example shows that even if the extrapolated absorption trend does not continue to 54 layer graphene, all is not lost.   
It uses the same 32 km transmitter and 1000 kg payload as the second example, but limits the number of layers to the experimentally measured range.
The final attainable velocity is decreased only slightly, from 0.149c to 0.119c.
A bigger drawback is that the roughly 3-fold lower acceleration means the time to final velocity is that much greater.
\section{Getting the data back}
Assuming we can launch a flyby probe, and it collects data as it passes through an alien solar system, we then have
the problem of how to get the data back.  Solar energy from the target star will be insufficient, as the probe is only within
any substantial illumination for at most a few hours.

The large transmitter provides a potential answer to this dilemma. The gain of a uniformly illuminated, coherent emission disk is
$$G = (2 \pi r / \lambda)^2$$
Multiplied by the large transmitted power, this results in a huge EIRP (roughly $5.20\cdot 10^{35}$ watts for the 32 km radius transmitter, 
compared to the roughly $3.85 \cdot 10^{26}$ watts of the sun).  This is enough to provide 4.6 w/m$^2$ at a distance of 10 light years.
Since this will arrive in monochromatic (though redshifted) form it should be especially efficient to convert this to useable
electrical energy.

Now to send the data back, assume a 1 meter aperture, and a 1 watt optical transmitter at $\lambda = 500$ nm.   Then the EIRP, 
computed as above,
is $3.95\cdot 10^{13}$ watts.  After 10 light years, this gives a received power density of $3.51\cdot 10^{-22}$ w/m$^2$.
Since each photon has $3.98\cdot 10^{-19}$ joules, a large collector is clearly needed to collect a significant number of photons.
Since we are assuming large, space-based structures are possible, perhaps assume a receiver about a quarter of the area of the
transmitter.  Such a 16 km radius telescope would collect about 700,000 photons per second.  
If we can pick them out of the background
light from the star, this should give a decent download rate.  The star (if sunlike) is roughly $10^{13}$ times brighter.
However we can apply very aggressive optical filtering (since the data light is monochromatic and of known wavelength),
temporal filtering (we could perhaps use 1 ps pulses every microsecond, for another gain of $10^6$), and coronograph techniques
currently being developed for the search for extrasolar planets.  Between these we should be able to distinguish the
data photons from the noise and get the probe's data back.
\section{Other uses}
Flyby missions for an Oort cloud object would be easy and cheap (no on-board power needed).

Such a system could make short work of any orbital debris, and deflect incoming asteroids or comets\cite{lubin2013directed}.  
Conversely, it would make a fine weapon system, dumping enormous power into a tiny area on any solar system body.
This possibility might argue for putting the transmitter outside the orbit of the Earth.  This would require a larger transmitter for the same output, but be physically unable to target the Earth.

Such a system could make a completely unambiguous way to make contact with an extraterrestial civilization, 
provided only the arriving wavelength is within the visible spectrum of the observer.  
An observer in the beam would see the parent star suddenly increase in brightness by $10^9$ times or so, 
until (for nearby stars) it casts an appreciable fraction of daylight 
(at 10 light years, it would be considerably brighter than typical office lighting here on Earth).  
Any society that watches the sky could not miss such a signal.
A message could be sent by blinking such a beam at a slow rate (perhaps 1 bit/sec) in a pattern that repeats with a period that is a product of two primes\cite{sagan1975recognition}.
Any society that has discovered prime numbers and factoring (such as the ancient Greeks or Chinese) might note that the 
repeat interval is the product of primes, say M x N, and then 
display the data on a M x N grid, and then attempt to decipher the message.
Such a system could therefore contact a civilization that is intelligent but not technological.
\section{Future work}
The required power, and transmitter size, could in theory be further reduced by propelling the sail with UV light, 
where the absorption of a single layer of graphene reaches 10\%\cite{kravets2010spectroscopic}.
However, there is currently no known method for generating UV light with high efficiency.
\section{Discussion}
Graphene sails and phased array optical drive can potentially reduce the scale of structures needed for interstellar probes.
No new physics is needed - ``only' the technologies of multi-acre graphene sheets, and kilometer scale nanotube ropes, and the ability to build large space-based structures. 
For comparison, the LHC is about 4.3 km in radius, the ISS about 420,000 kg, 
and large photovoltaic power plants are perhaps 5 km$^2$ in size and more than 0.5 GW in output.
These are within 1-2 orders of magnitude of what is needed.
Furthermore, most of the technologies needed (efficient solar cells, efficient lasers, precision space constellations, graphene and nanotube production) are being developed for other reasons.

For such a system, the vast majority of the cost is in the transmitter.  The probes are themselves should be relatively cheap.
Furthermore, the launch time is relatively short, then there is a long coast where the time depends on the destination, 
then a relatively quick data retrieval.  
So this architecture makes sense for launching, then retrieving data from a fleet of probes.
\section{Conclusion}
This paper continues the work of investigating different possible architectures for interstellar probes, 
searching for those that may be more economical and practical.

Even with the possible improvements outlined here,
the facilites for interstellar probes will still be large, and too expensive for the current earth economy.  
They will not be built in the next few years, nor in the next few decades, but perhaps they are not so far out of reach as has been thought.
\section*{Appendix 1: Code for calculations}
Here is a program in C++ that performs the calculations used in the table.  The input is a file that contains examples,
one per line, with the transmitter radius in meters, the maximum number of layers in the sail, the areal power output of the transmitter, 
and the payload mass.
The output is a table in LaTex format, called ``tab.tex''.

This program does not use the analytic
results from the paper concerning the number of layers or the optimum payload/sail mass ratio.  Instead the
optimum is found by gradient descent.
{\tiny
\begin{lstlisting}[language=C]
#include <stdio.h>
#include <math.h>
#include <vector>
using namespace std;

const double GRAPHENE_ABSORBS = 0.0230;   // one layer absorbs 2.3% of visible light
const double GRAPHENE_MASS    = 0.77e-6;  // mass of 0.77 mg/m^2
const double GRAPHENE_THICK   = 0.345e-9; // thickness of a layer
const double GRAPHENE_STRENGTH= 1.0e11;   // Pascals, graphene
const double RIGGING_STRENGTH = 100e9;    // Pascals, carbon nanotube
const double RIGGING_DENSITY  = 1300;     // kg/m^3, carbon nanotubes
const double SI_DENSITY       = 2329;     // kg/m^3, silicon
const double GaAs_DENSITY     = 5320;     // kg/m^2, Gallium Arsenide

const double STEFAN_BOLTZMANN = 5.67e-8;  // Stefan-Boltzmann constant
const double C                = 3e8;      // speed of light, m/s
const double G                = 9.8;      // g at Earth surface m/s^2
const double AU               = 1.496e11; // AU in meters

const double LAMBDA            = 500e-9;  // 500 nm light
const double LENS_THICKNESS    = 10e-6;   // 10 microns thick
const double ISS_MASS          = 419455;  // mass of international space station in kg

class print {
  public:
    const char *label;
    const char *format;
    double scale;
    bool line;
    print(const char *a, const char *b = "%.1f", double s = 1.0, double l = false){label = a; format = b; scale = s; line = l;}
};

// calculate one specific case
double one_calc(double tr, double sr, int nl, double pmass, double power_per_area, vector<double> &rslt) {
rslt[0] = tr;
rslt[6] = sr;
rslt[8] = nl;
rslt[14] = pmass;
printf("\nTrans radius %.1f, sail rad %.1f, %d layers, payload %.3f\n", tr, sr, nl, pmass);
double tr_area = rslt[1] = M_PI *pow(tr,2);
rslt[2] = power_per_area;
double power =   rslt[3] = tr_area * power_per_area;
printf("area of transmitter %.1f km^2, power %.1f gigawatts\n", tr_area/1e6, power/1e9);
double tr_mass = rslt[4] = rslt[5] = tr_area* LENS_THICKNESS*GaAs_DENSITY;
printf("Mass of lens %.1f kg, %.2f space stations\n", tr_mass, tr_mass/ISS_MASS);
double absorb =  rslt[10] = 1 - pow(1-GRAPHENE_ABSORBS, nl);
printf("A sail with %d layers absorbs %.2f%% of incident light.\n", nl, absorb*100);
double sail_area  = rslt[7] = M_PI * pow(sr,2);
double sail_power_per_area = rslt[11] = power * absorb /sail_area;
double sail_mass = rslt[9] = sail_area * GRAPHENE_MASS * nl;
double temp = rslt[12] = pow(sail_power_per_area/(2*STEFAN_BOLTZMANN*absorb), 0.25);
printf("Sail area %.1f km^2, power density %.1f w/m^2, temp %.1f K, mass %.3f kg\n", 
 sail_area/1e6, sail_power_per_area, temp, sail_mass);
double thrust = rslt[13] = sail_power_per_area * sail_area / C; 
double acc =  rslt[15] = thrust/(pmass + sail_mass);
printf("Initial guess at acceleration %.3f m/s, %.3f gs\n", acc, acc/G);
double rigging_length = sqrt(2) * sr;
double rigging_cross_area = sqrt(2) * pmass*acc /RIGGING_STRENGTH;
double rigging_mass = rslt[16] = RIGGING_DENSITY * rigging_length * rigging_cross_area;
printf("Mass of rigging %.3f kg\n", rigging_mass);
acc = rslt[17] = thrust/(pmass + sail_mass + rigging_mass);
printf("Better guess at acceleration %.3f m/s, %.3f gs\n", acc, acc/G);
rslt[18] = LAMBDA;
double dist = rslt[19] = 4 * sr * tr/LAMBDA/2.44;
printf("Distance until can't focus %.1f km, %.2f AU\n", dist, dist/AU);
double v = rslt[20] = rslt[21] = sqrt(2 * acc * dist);
double t = rslt[22] = sqrt(2 * dist / acc);
printf("Newtonian v=%.1f m/s, %.4f c;  t=%.1f sec, %.2f days %.3f years\n", v, v/C, t, t/86400, t/365.25/86400);
double err = 10;
do {
    double eterm = -expm1(-acc/C*t);
    v = C * eterm;
    double s = C*t - C*C/acc*eterm;
    err = dist-s;
    t += err/v;
    printf("Relativistic s=%.1f v=%.1f err=%.3f\n", s, v, err);
    }
while (fabs(err) > 100.0);
rslt[23] = v;
rslt[24] = t;
printf("Relativistic v=%.1f m/s, %.4f c;  t=%.1f sec, %.2f days %.3f years\n", v, v/C, t, t/86400, t/365.25/86400);
double thick = nl*GRAPHENE_THICK;
double hoop  = GRAPHENE_STRENGTH * 6.0/13.0;
double edge  = rslt[25] = thrust/(2*M_PI)/thick/hoop;
printf("Requires outer %.3f meters to make hoop stress positive\n", edge);
return v/C;
}

int main(int argc, char **argv) {
printf("enter radius of transmitter (meters), radius of sail (meters), number of monolayers, mass of payload (kg):\n");
vector<double> trad, srad, pm, ppa;
vector<int> layers;
for(;;) {
    double tr, sr, pmass, power_per_area;
    int nlmax;
    if (scanf("%lf %d %lf %lf", &tr, &nlmax, &pmass, &power_per_area) != 4)
	break;
    printf("Got %f\n", tr);
    trad.push_back(tr);
    srad.push_back(tr/10.0);  // start with sail 1/10 transmitter size
    layers.push_back(nlmax);  // max number of layers
    pm.push_back(pmass);
    ppa.push_back(power_per_area);
    }

vector<vector<double> > table;
for(int i=0; i<trad.size(); i++) {
    int LMAX = layers[i];
    layers[i] = 1;
    // loop until find the right value
    vector<double> numbers(26);
    for(;;) {
	vector<double> rslt;
	double delta = 0.001 * srad[i];
	for(int n=layers[i]-1; n <= layers[i]+1; n++) {
	    for (double r = srad[i] - delta; r <srad[i]+1.5*delta; r += delta) 
	       rslt.push_back(one_calc(trad[i], r, n, pm[i], ppa[i], numbers));
	    }
        if (layers[i]+1 > LMAX)
	    rslt[6] = rslt[7] = rslt[8] = 0.0;
	printf("  rad=%9.1f %9.1f %9.1f\n", srad[i]-delta, srad[i], srad[i]+delta);
	for(int n=0; n<rslt.size(); n+=3)
	    printf("n=%3d %9.5f %9.5f %9.5f\n", layers[i] + (n/3-1), rslt[n], rslt[n+1], rslt[n+2]);
	//find the biggest one.  If it's not number 5, readjust and try again.
	double biggest = 0.0;
	int bigi = -1;
	for(int k=0; k<rslt.size(); k++) {
	    if (rslt[k] > biggest) {
		biggest = rslt[k];
		bigi = k;
		}
	    }
	printf("Biggest is case %d\n", bigi);
	if (bigi == 4)
	    break;
	int ncase = bigi/3;
	layers[i] += (ncase-1);
	int rcase = (bigi % 3);
	srad[i] += (rcase - 1) * delta;
	}
    // OK, now (re) compute final result
    one_calc(trad[i], srad[i], layers[i], pm[i], ppa[i], numbers);
    table.push_back(numbers);
    }
print labels[] = {print("Radius of transmitter (km)", "%.2f", 1e-3),               // [0]
                      print("Area (square km)", "%.1f", 1e-6),                     // [1]
                      print("Transmitter flux (w/m$^2$)", "%.1f", 1.0),
                      print("Power (GW)", "%.1f", 1e-9),                    // [3]
                      print("Mass of transmitter (tonnes)", "%.1f", 1e-3),                     // [6]
                      print("Mass in units of ISS", "%.1f", 1.0/ISS_MASS, true),    // [7]
                      print("Radius of sail (meters)"),           // [3]
                      print("Sail area (km$^2$)", "%.3f", 1e-6),    // [9]
                      print("Number of layers"),         // [4]
                      print("Sail mass (kg)"),    // [10]
                      print("Absorbence (\\%)", "%.2f", 100.0),            // [8]
                      print("Sail absorb kW/m$^2$", "%.1f", 1e-3),    // [11]
                      print("Sail temperature (Kelvin)"),    // [12]
                      print("Thrust (newtons)", "%.3f", 1.0, true), // [13]
                      print("Payload mass (kg)", "%.3f", 1.0),             // [5]
                      print("Acceleration, sail+payload only(m/s$^2$)"),     // [14]
		      print("Rigging mass (kg)", "%.3f"),  /// [15]
                      print("Acceleration including rigging(m/s$^2$)"), // [16]
                      print("Wavelength of light used (nm)", "%.1f", 1e9), // [17]
                      print("Distance to cutoff (AU)", "%.1f", 1.0/AU), // [18]
                      print("V at cutoff, Newtonian (km/sec)", "%.1f", 1e-3),                  // 19
                      print("Newtonian V (in term of C)", "%.3f", 1.0/C),    // 20
                      print("Newtonian time (days)", "%.1f", 1.0/86400.0),          // 21
                      print("Relativistic V (in term of C)", "%.3f", 1.0/C), // 22
                      print("Relativistic time (days)", "%.1f", 1.0/86400.0),      // 23
		      print("Rim width (meters)", "%.3f")};               // 24
// write table in tex format
FILE *ftab = fopen("tab.tex","w");
if (ftab == NULL) {
   printf("can't open tex file\n");
   return 42;
   }
const char *cs = (argc > 1) ? "%" : "";  // comment string
fprintf(ftab, "%s\\documentclass[10pt]{article}\n"
              "%s\\usepackage{array}\n"
              "%s\\usepackage[margin=0.5in]{geometry}\n"
              "%s\\begin{document}\n", cs,cs,cs,cs);
fprintf(ftab, "%s\\begin{tabular}{ r", cs);
for(int i=0; i<table.size(); i++)
     fprintf(ftab, " r");
fprintf(ftab, "}\n");
fprintf(ftab, "%sExample ID ", cs);
for(int i=0; i<table.size(); i++)
     fprintf(ftab, "& (%c)", 'A' + i);
fprintf(ftab, "\\\\ \\hline\n");

for(int i=0; i<=24; i++) {
    fprintf(ftab, "%s ", labels[i].label);
    for(int k=0; k<table.size(); k++) {
	fprintf(ftab, "& ");
	fprintf(ftab, labels[i].format, table[k][i]*labels[i].scale );
        }
    fprintf(ftab, "\\\\%s\n", labels[i].line ? "\\hline" : "");
    }

fprintf(ftab, "\\end{tabular}\n");
fprintf(ftab, "%s\\end{document}\n", cs);
fclose(ftab);
return 0;

}
\end{lstlisting}
}
\bibliography{graphene}

\end{document}

%% file: tab.tex
%\documentclass[10pt]{article}
%\usepackage{array}
%\usepackage[margin=0.5in]{geometry}
%\begin{document}
%\begin{tabular}{ r r r r r r}
%Example ID & (A)& (B)& (C)& (D)& (E)\\ \hline
Radius of transmitter (km) & 7.65& 32.00& 32.00& 80.00& 32.00\\
Area (square km) & 183.9& 3217.0& 3217.0& 20106.2& 3217.0\\
Transmitter flux (w/m$^2$) & 1000.0& 1000.0& 1000.0& 1000.0& 1000.0\\
Power (GW) & 183.9& 3217.0& 3217.0& 20106.2& 3217.0\\
Mass of transmitter (tonnes) & 9781.0& 171143.9& 171143.9& 1069649.5& 171143.9\\
Mass in units of ISS & 23.3& 408.0& 408.0& 2550.1& 408.0\\\hline
Radius of sail (meters) & 87.5& 2767.8& 175.3& 2770.2& 7190.4\\
Sail area (km$^2$) & 0.024& 24.066& 0.097& 24.108& 162.426\\
Number of layers & 54.0& 54.0& 54.0& 54.0& 8.0\\
Sail mass (kg) & 1.0& 1000.7& 4.0& 1002.4& 1000.5\\
Absorbence (\%) & 71.54& 71.54& 71.54& 71.54& 16.99\\
Sail absorb kW/m$^2$ & 5466.3& 95.6& 23839.4& 596.6& 3.4\\
Sail temperature (Kelvin) & 2865.1& 1042.0& 4140.4& 1646.8& 646.5\\
Thrust (newtons) & 438.401& 7670.948& 7670.948& 47943.426& 1821.354\\\hline
Payload mass (kg) & 1.000& 1000.000& 4.000& 1000.000& 1000.000\\
Acceleration, sail+payload only(m/s$^2$) & 219.2& 3.8& 957.2& 23.9& 0.9\\
Rigging mass (kg) & 0.000& 0.276& 0.017& 1.724& 0.170\\
Acceleration including rigging(m/s$^2$) & 219.1& 3.8& 955.1& 23.9& 0.9\\
Wavelength of light used (nm) & 500.0& 500.0& 500.0& 500.0& 500.0\\
Distance to cutoff (AU) & 14.7& 1941.1& 122.9& 4857.0& 5042.8\\
V at cutoff, Newtonian (km/sec) & 31013.7& 47185.8& 187436.1& 186450.7& 37061.3\\
Newtonian V (in term of C) & 0.103& 0.157& 0.625& 0.622& 0.124\\
Newtonian time (days) & 1.6& 142.5& 2.3& 90.2& 471.2\\
Relativistic V (in term of C) & 0.100& 0.149& 0.502& 0.500& 0.119\\
Relativistic time (days) & 1.7& 146.3& 2.5& 100.6& 481.1\\
\end{tabular}
%\end{document}